# SafeEmail – A safe and reliable email communication system without any spam

Avanna　　　Psycho Zhang　　　Simon Yan　　　Jonne Mao

*Version: 0.1.0*

*February 12, 2019*

**Abstract**

Using multi-group asymmetric public and private keys, this paper proposes a encryption e-mail communication system, which makes e-mail communication more secure, lowers the service provider's network and storage consumption, and completely eliminates any spam.

Prior to this, PGP (Pretty Good Privacy) partially solved the encryption problem, but it only encrypts the content of the mail. We would also expect a mail system to provide safe and reliable services, completely eliminate the spam emails, reduce the consumption of service providers' traffic and storage resources, and specially focus on protecting the user privacy and the data security of service provider.

## 1　Foreword

When you use Internet services, all your actions and traces are exposed to the risk of being analyzed or sold. As long as there is a possibility of obtaining data, your personal information and data will be stolen by various methods, and this does not require your consent or permission.

We are not opposed to the fact that the true behavioral trajectory is recorded, but we do not want this information to become available for analysis or to be sold, and we do not want our identity to be accurately used by criminals.

Maybe the probability of big data error is only one in 10 million, maybe the probability of data being stolen is only one in 10 million, and the probability that the database containing your personal information is maliciously broke is only one in 10 million, but once it happens, it might cause significant losses to you.

## 2　Introduction

The behavioral trajectory of each of us on the Internet has been preserved and recorded by countless websites, applications, and systems, unconsciously. And every piece of data may be tampered with or forged by any third party;

What's more, by putting together these various behavioral trajectories, someone can perform accurate analysis, your behavior, your preferences, your taste, height and weight, your phone number, home address, and car model, even where your children's schools are... and so on, everything will be exposed. When the data is being sold, you have no secret in the eyes of the data buyers.

In an existing email communication system, you may encounter the following: Your friend or a family member received a message from you. All the evidence indicates that the email is authentic. However, after the family pay the money according to the information provided by the email, it was discovered that you did not send this email. Your password has not been lost yet; A virus outbreaks, a malicious extortion email was shown to be issued by you, the blackmailed company reported the case, and the police knocked at your front door; After you sent your cover letter to your favorite company, you soon received a reply for the interview. And you didn't realize that this reply is forged. When you went to the appointment happily, it turned out that you were cheated by an illegal organization; etc.

In real life, such examples are everywhere.

Although there are now mail systems that support the encryption of letter content, this does not prevent counterfeiting and tampering. A fake email can be encrypted and sent to you. Therefore, encryption does not solve the problem of authenticity.

Therefore, we believe that people need some kind of e-mail or communication system, which is based on the principle of cryptography. All email accounts and contents cannot be changed, or forged, and could be deliver efficiently.

This paper aims to propose such a solution.

## 3  Expectations

### 3.1  Protection Of Information Data Integrity

In the existing e-mail system, the user's account number and password can be registered and used after the service provider confirms that it is available. In other words, the user assumes that the account password is set by itself, but is actually assigned to the user by the service provider - so the service provider has the ability to forge user information.

We propose an authorization mechanism for checks and balances between users and service providers, that is, the user's account and password are generated by the user himself, but the service must be used only after obtaining the authorization of the service provider.

Under the traditional account password system, the process of using the service is: The user submits an account and password to the service provider, and after the service provider allows, the user starts to use the service.

Under the system of authorities and balances, the process of using services is: The user submits his signature of the service to the service provider, and after the service provider verifies that the signature matches, the service is started.

The difference between the two systems is: Under the traditional account password system, because the service provider holds the user's account password, it has the ability to forge user behavior. When a hacker obtains the user's account password, it can also falsify user behavior. When a service provider is subjected to a malicious attack or is doing something evil, the user might be responsible for the adverse consequences of the behavior that he didn't have before.

Under the system of authorities and balances, the service provider has no ability to falsify user behavior. So even if the service provider is maliciously attacked, the attacker has no ability to falsify the user's signature and then falsify the user's behavior.

The so-called information integrity, in our view, in addition to ensuring the integrity of the content, it is also necessary to ensure the integrity of the source information.

### 3.2 Multi-center Communication And Privacy Data Protection

User profile information, including user account names and behavioral trajectories, are among the privacy data categories of service providers and users. In the existing mail communication system, in order to provide services, service providers have to exchange user information with other service providers. For example, the mail service provider must provide user inquiry functions to other mail service providers so that the mail can be correctly delivered.

There is a potential danger of potentially revealing user privacy. For example, a malicious attacker can initiate a collision attack by querying the existence of an account of a certain person at a service provider, thereby causing a series of information leakage of the user. This security responsibility lies entirely with the user, since he has set up the same account and password at different service providers.

We propose a new black box service system that uses a zero-knowledge proof method to enable service providers to provide email communication services to users without any user privacy information.

### 3.3 Cannot Selectively Maliciously Delay Or Refuse Service

Under the existing mail communication system, the user's information delivery target is transparent to the service provider. After collecting the information interaction time and interaction frequency, the service provider or malicious attacker can analyze the user behavior in addition to the user behavior. They can maliciously and selectively delay or even block your interaction with the target to achieve their desired goals. And you may think that it is only caused by some other uncontrollable factors.

We want that when users are sending emails, it is also a black box service model based on zero-knowledge proof, even for service providers. The service provider can record the user's behavior track, and the malicious attacker also has the possibility of intercepting the data, but the data cannot be analyzed or used without user authorization.

# 4 Realization

## 4.1 Service Contract

Assumed: User Alice registers a mail communication service at the service provider Gogo.

In order to prevent spam, the addresses of the senders who will deliver mails to Alice's inbox must be authorized by Alice, otherwise the mails will be filtered and discarded.

In other words, only the address authorized by Alice's signature has the right to submit mail information to Alice.

A user will generate two pairs of keys, one pair is major, used for service provider authentication. The other pair is minor, which is used to authorize other people to leave a message.

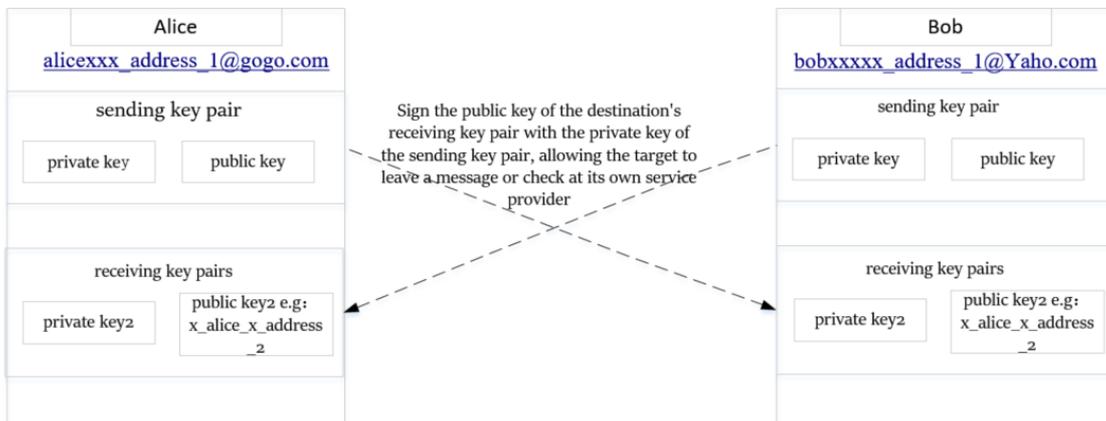

## 4.2 Authorizing Sender Address

For example, when Bob sends a message to Alice, he should tell Alice his own sending address in advance. Only after Alice signs Bob's sender address with a private key and notifies the service provider, Bob can send mail to Alice's inbox.

It should be noted here that the inbox space provided by the service provider Gogo to Alice could be very small, and the maximum amount of information delivered to the inbox is no more than 1 KB.

Assumed: Alice's email address is: alicexxx_address_1@gogo.com, which is approved by Gogo. The sender addresses authorized by alicexxx_address_1 is also approved by Gogo and can send no more than 1K information to the Alice inbox. At the same time, Alice also has another private keys whose public key is: x_alice_x_address_2, which will only be exposed to other users who will communicate with her.

Same as above, Bob registers a safe mail service at the service provider Yaho. Bob's email address is: bobxxxxx_address_1@gogo.com, which is approved by Yaho. The public key of the Bob's another private key is: x_bob_x_address_2.

### 4.3 Sending Mail

So how does Bob, who uses the Yaho service, send an email to Alice using the Gogo service?

#### 4.3.1 Upload Mail Content

Bob will package the mail content and Alice's public key address x_alice_x_address_2 and the randomly generated extraction code, for example: 123456, and send it to the service provider Yaho. And tell them that my e-mail should be sent to the public key address x_alice_x_address_2, if someone can sign the information of the extraction code 123456, then you can let him take the mail.

It should be noted that Yaho does not know who x_alice_x_address_2 is. Anyone who wants to view this email have to verify the signature. i.e. Verify("123456", "x_alice_x_address_2", signature) If it matches, they could access the message.

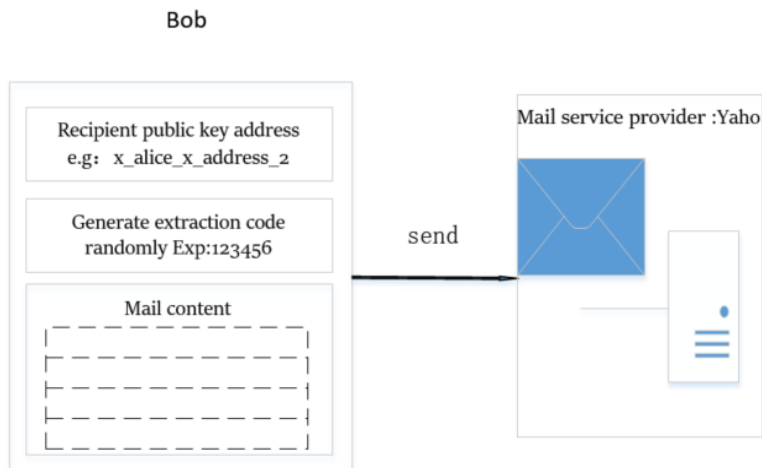

Bob's mail service provider learned that Bob had an email sent to a public key address. Although it does not know who the public key address belongs to, it can verify the signature of the extraction code. The correct signature must be the legal one.

#### 4.3.2 Leave A Message

Bob uses the authorized account signed by Alice to log in to the service provider Gogo, and sends the code 123456 to Alice.

It should be noted that Bob logs into the Gogo server and uses the address that Alice has authorized to sign x_bob_x_address_2, so the Gogo server does not know the real identity of Bob.

Now Bob finished sending mail to Alice.

Bob submitted the mail content on Yaho's server and sent a message to Alice on the Gogo server. Then just wait for Alice to get the notification and she will get the mail content from Yaho's server.

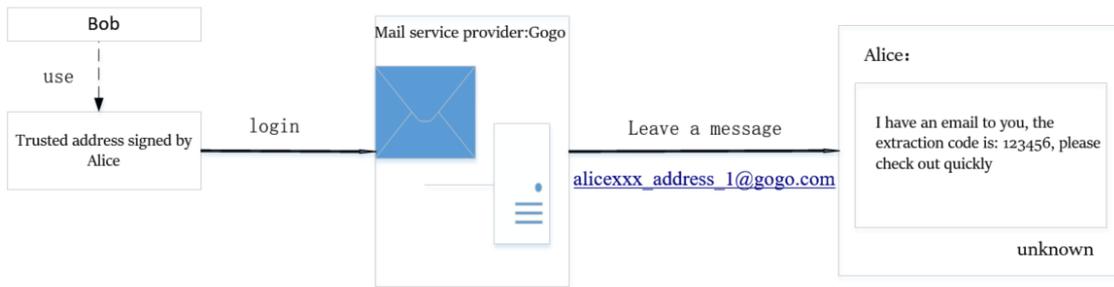

Bob leaves Alice a message at Alice's mail service provider

## 4.4 Receiving Mail

### 4.4.1 Check The Message

Alice logs in to his Gogo mailbox, checks the message, and gets the extraction code left by Bob.

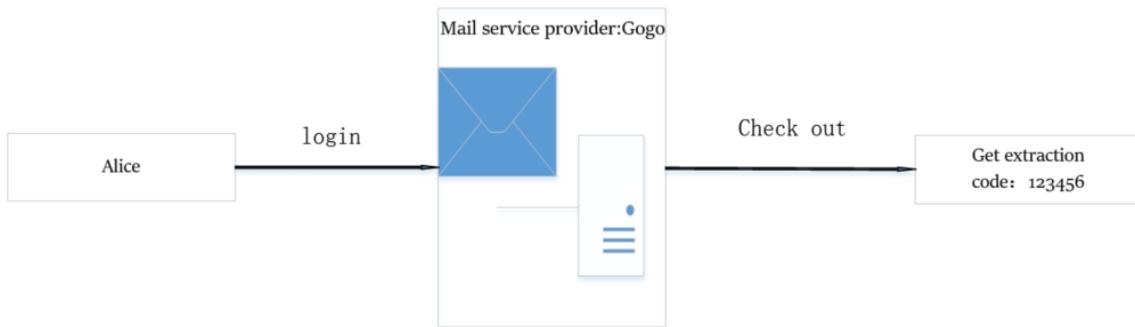

Alice logs in to his email service provider to get Bob's message to herself and get the extraction code.

### 4.4.2 Read Mail Content

Alice receives the Bob message extraction code, and uses the private key of x_alice_x_address_2 to sign the extraction code. The signature and the address of the pickup are verified by the service provider Yaho and can be retrieved.

It should be noted that Alice knows who x_bob_x_address_2 is and which service provider should go to get real email information.

So Alice could pickup the mail successfully.

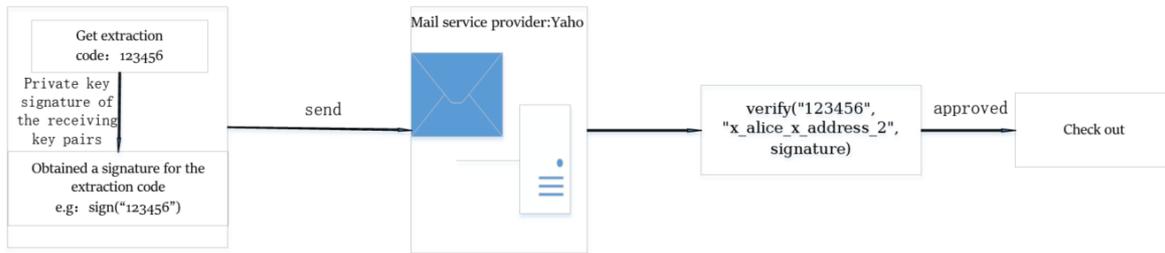

Alice sends signature of the extraction code to the Bob's mail service provider for verification by the private key in the own receiving key pairs.

# 5  Summary

Only information sent from the sender's address authorized by the user's signature is allowed to be placed in the inbox, thus completely eliminating spam.

## 5.1  Put An End To Spam

Bob's mail to Alice is not stored on Gogo's server, so Gogo's storage consumption is zero. Yaho does not send mail content to Gogo over the network, so Yaho's network traffic consumption is zero. Storage and traffic consumption are reduced by 50% compared to existing mail communication systems.

## 5.2  Reduced Service Provider Storage Space And Network Traffic Consumption

Bob's mail to Alice is not stored on Gogo's server, so Gogo's storage consumption is zero. Yaho does not send mail content to Gogo over the network, so Yaho's network traffic consumption is zero. Storage and traffic consumption are reduced by 50% compared to existing mail communication systems.

## 5.3  Privacy Protection

During Bob's entire mailing process, the service provider Yaho recorded the behavior of Bob's mail message to the mail, and also saved the signature of the pickup, providing Bob with a complete service. From the perspective of these two transactions, there is no privacy exposed.

In the process of Alice's pickup, the service provider Gogo provided Alice with a service, and Gogo learned and recorded the behavior of someone posting a message to Alice. And this independent transaction is also no privacy exposed.

## 5.4  Data Integrity

Neither Gogo nor Yaho have the users' private keys. Information must be signed and authenticated from the beginning to the end, so neither Gogo nor Yaho, or a malicious attacker, can tamper with or falsify. At the same time, because the service provider does not know the identity information of the recipient, it cannot even selectively delay or deny the service.

## 5.5 Implementation

Based on the combination of zero-knowledge proof and asymmetric encryption, a privacy-protected, secure, spam-free mail communication system is implemented. The implementation is open source on github (https://github.com/maikejonne), including server implementations and client-side demo.

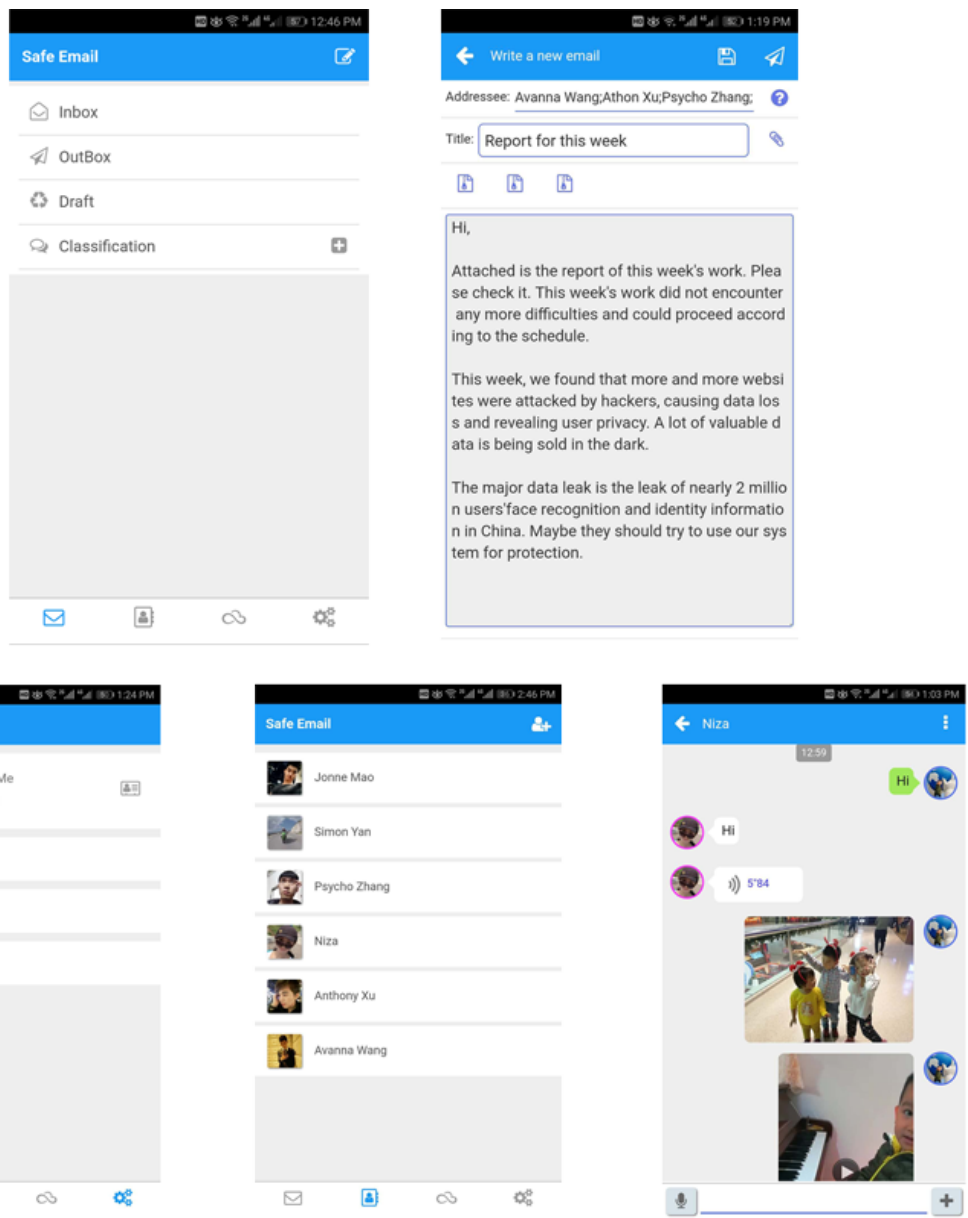

These images are taken from the project